\documentclass[
    aps,
    prd,
    nofootinbib,
    superscriptaddress,
    10pt,
    notitlepage,
    twocolumn]{revtex4-1}

\usepackage{anyfontsize}
\usepackage{amssymb}
\usepackage{amsmath}
\usepackage{amsfonts}
\usepackage{amsbsy}

\usepackage{newtxtext}
\usepackage{newtxmath}

\usepackage{tensor}
\usepackage{mathrsfs}
\usepackage{bm}

\usepackage[utf8]{inputenc}

\usepackage{graphicx}
\usepackage{epsfig}
\usepackage{epstopdf}

\usepackage[linktocpage,breaklinks]{hyperref}
\usepackage[usenames,dvipsnames]{xcolor}

\definecolor{romared}{RGB}{142,0,28}
\hypersetup{colorlinks=true,
            citecolor=romared,
            linkcolor=romared,
            urlcolor=romared}

\newcommand{\nn}{\nonumber}

\newcommand{\be}{\begin{equation}}
\newcommand{\ee}{\end{equation}}

\begin{document}

\title{Ultralight boson cloud depletion in binary systems}

\begin{abstract}
  Ultralight scalars can extract rotational energy from astrophysical
  black holes through superradiant instabilities, forming macroscopic
  boson clouds. This process is most efficient when the Compton
  wavelength of the boson is comparable to the size of the black hole
  horizon, i.e. when the ``gravitational fine structure constant''
  $\alpha\equiv G \mu M/\hbar c\sim 1$.  If the black hole/cloud
  system is in a binary, tidal perturbations from the companion can
  produce resonant transitions between the energy levels of the cloud,
  depleting it by an amount that depends on the nature of the
  transition and on the parameters of the binary.  Previous cloud
  depletion estimates considered binaries in circular orbit and made
  the approximation $\alpha\ll 1$. Here we use black hole perturbation
  theory to compute instability rates and decay widths for generic
  values of $\alpha$, and we show that this leads to much larger cloud
  depletion estimates when $\alpha \gtrsim 0.1$. We also study
  eccentric binary orbits. We show that in this case resonances can
  occur at all harmonics of the orbital frequency, significantly
  extending the range of frequencies where cloud depletion may be
  observable with gravitational wave interferometers.
\end{abstract}

\author{Emanuele Berti}
\email{berti@jhu.edu}
\affiliation{Department of Physics and Astronomy,
Johns Hopkins University, 3400 N. Charles Street, Baltimore, MD 21218, USA}

\author{Richard Brito}
\email{richard.brito@roma1.infn.it}
\affiliation{Dipartimento di Fisica, ``Sapienza'' Universit\`a di Roma \& Sezione INFN Roma1, Piazzale Aldo Moro 5, 00185, Roma, Italy}

\author{Caio F. B. Macedo}
\email{caiomacedo@ufpa.br}
\affiliation{Campus Salin\'opolis, Universidade Federal do Par\'a,
Salin\'opolis, Par\'a, 68721-000 Brazil}

\author{Guilherme Raposo}
\email{guilherme.raposo@roma1.infn.it}
\affiliation{Dipartimento di Fisica, ``Sapienza'' Universit\`a di Roma \& Sezione INFN Roma1, Piazzale Aldo Moro 5, 00185, Roma, Italy}

\author{Jo\~ao Lu\'is Rosa}
\email{joaoluis92@gmail.com}
\affiliation{Department of Physics and Astronomy,
Johns Hopkins University, 3400 N. Charles Street, Baltimore, MD 21218, USA}
\affiliation{Centro de Astrof\'isica e Gravita\'c\~ao - CENTRA,
Departamento de F\'isica, Instituto Superior T\'ecnico - IST,
Universidade de Lisboa - UL, Avenida Rovisco Pais 1, 1049-001, Portugal}

\date{{\today}}
\maketitle

\section{Introduction}
\label{sec:int}

The observation of gravitational waves (GWs) by the LIGO and Virgo
collaborations~\cite{LIGOScientific:2018mvr} marked the beginning of a
new era in astrophysics and fundamental
physics~\cite{Barack:2018yly}. These observations have already
provided crucial information on the formation of binary compact
objects~\cite{LIGOScientific:2018jsj}, tested general relativity in
the strong, highly dynamical
regime~\cite{TheLIGOScientific:2016src,Yunes:2016jcc,Berti:2015itd,Berti:2018cxi,Berti:2018vdi},
and led to new measurements of the cosmological expansion of the
universe~\cite{Abbott:2017xzu,Soares-Santos:2019irc}.

Quite remarkably, GW observations also have the potential to transform
our understanding of particle physics. One example is the possibility
to discover ultralight bosonic particles, such as axions, through
GWs~\cite{Arvanitaki:2014wva,Yoshino:2014wwa,Arvanitaki:2016qwi,Baryakhtar:2017ngi,Brito:2017wnc,Brito:2017zvb,DAntonio:2018sff,Isi:2018pzk,Ghosh:2018gaw,Tsukada:2018mbp}. Ultralight
bosons can efficiently extract rotational energy from spinning black
holes (BHs) through superradiant instabilities and form macroscopic
condensates when the Compton wavelength of the boson is comparable to
the characteristic size of the BH horizon, i.e. when the
``gravitational fine structure constant''
$\alpha\equiv G \mu M/\hbar c\sim 1$~\cite{Brito:2015oca}.  This
possibility can shed light on bosons with masses in the range
$\sim[10^{-19},10^{-11}]$ eV, which have Compton wavelengths
comparable to the size of astrophysical BHs.

The existence and formation of bosonic clouds can be inferred through
several observational channels. A first possibility is to look for
gaps in the ``Regge plane'' of astrophysical BHs: superradiant
instabilities could lead to a lack of highly spinning BHs in a BH mass
range that depends on the boson mass. Measurements of the spin and
mass of astrophysical BHs can then be used to infer or constrain the
existence of ultralight
bosons~\cite{Arvanitaki:2014wva,Brito:2014wla,Baryakhtar:2017ngi,Cardoso:2018tly,Stott:2018opm}.
Even more exciting is the prospect of direct detection: once formed,
boson clouds would slowly decay through the emission of long-lived,
nearly monochromatic GWs. This radiation is potentially observable,
either as a continuous, nearly monochromatic signal from individual
sources or as a stochastic
background~\cite{Arvanitaki:2014wva,Arvanitaki:2016qwi,Baryakhtar:2017ngi,Brito:2017wnc,Brito:2017zvb,Ghosh:2018gaw,Tsukada:2018mbp}.

Here we study how bosonic clouds around astrophysical BHs can affect
the dynamics of a binary system, revisiting and extending the recent,
comprehensive analysis of~\cite{Baumann:2018vus}. The cloud can affect
the motion of small compact objects in its
vicinity~\cite{Ferreira:2017pth,Zhang:2018kib}. The special nature of
the axionic cloud would leave characteristic signatures in the
gravitational waveforms from extreme mass-ratio inspirals, that are
potentially detectable by LISA~\cite{Hannuksela:2018izj}.
In this work we consider the effect of a binary companion on the
bosonic cloud itself~\cite{Arvanitaki:2014wva}.  Under certain
conditions, the perturbations induced by the companion can lead to
resonant transitions between superradiant and non-superradiant modes,
which can, in some cases, deplete the cloud~\cite{Baumann:2018vus}.
Ref.~\cite{Baumann:2018vus} assumed that $\alpha\ll 1$ and that the
binary is on a circular, equatorial orbit that can be treated in the
Newtonian limit.
We relax the approximation $\alpha\ll 1$ (which is expected to fail
precisely when the superradiant instability is strongest) by using
numerical calculations in BH perturbation theory to estimate the decay
rate of the non-superradiant modes, and we point out a sign error that
affects the decay rates of Ref.~\cite{Baumann:2018vus}.  We also
extend their analysis to eccentric binaries, showing that multiple
resonant depletion episodes can occur for eccentric BH binaries of
interest for LISA.

The plan of the paper is as follows. In Section~\ref{sec:cloud} we
review the hydrogenic structure of the energy levels of boson clouds
around rotating BHs. In Section~\ref{sec:resonances} we discuss
resonances in circular and eccentric binary systems, and in
Section~\ref{sec:results} we present our estimates for cloud
depletion. In Section~\ref{sec:conclusion} we highlight some
limitations of our study and point out directions for future work.  To
improve readability and to keep this paper self-contained, we relegate
some necessary technicalities to the
Appendices. Appendix~\ref{app:continuedfrac} shows
(following~\cite{Dolan:2007mj}) how to compute instability rates and
decay widths for generic values of $\alpha$ using continued fraction
methods. Appendix~\ref{app:levelmixing} deals with level mixing
induced by tidal perturbations and with the resulting selection rules,
summarizing some important results
from~\cite{Baumann:2018vus}. Finally, Appendix~\ref{app:ecc} presents
an approximate analytical calculation of the occupation numbers of
decaying levels valid for small-eccentricity orbits.

\section{Hydrogenic structure of the boson cloud}
\label{sec:cloud}

Superradiant instabilities can lead to the formation of ultralight
boson clouds around rotating (Kerr) BHs.  Consider a scalar field
$\Psi$ of mass $\mu$, described by the Klein-Gordon equation on a Kerr
background:
\begin{equation}\label{scalareom}
\left(\Box-\mu^2\right)\Psi(t,\textbf{r})=0,
\end{equation}
where $\Box=g^{ab}\nabla_a\nabla_b$ is the d'Alembert operator,
$g^{ab}$ is the contravariant Kerr metric, and $\nabla_a$ denotes a
covariant derivative. The angular dependence of the scalar field can
be separated with the ansatz 
\begin{equation}\label{scalarsep}
\Psi=\sum_{\ell,m} e^{i m\phi}e^{-i\omega t}S_{\ell m}(\theta) \psi_{n\ell m} (r)\,,
\end{equation}
which leads to ordinary differential equations for the radial
eigenfunctions $\psi_{n\ell m}(r)$, where $n$ is an integer labeling
the discrete eigenfrequencies $\omega$.
Modes with angular frequency
$\omega$ and azimuthal number $m$ will be superradiantly amplified if
the BH rotates faster than the field's phase velocity, i.e.
\begin{equation}\label{superradiance}
0<\omega<m\Omega_H\,,
\end{equation}
where $\Omega_H=\frac{a}{2M r_+}$ is a function of the Kerr angular
momentum parameter $a=J/M$ (where $M$ and $J$ are the BH mass and
angular momentum), and $r_+\equiv M+\sqrt{M^2-a^2}$ denotes the
Boyer-Lindquist horizon radius (here and below we will use geometrical
units, $G=c=1$). Let us also define $\tilde{a}=a/M$ and
$\tilde{r}_+\equiv r_+/M$ for future use.
The mass of the scalar field works as a potential barrier that
confines the superradiant modes, leading to a continuous extraction of
angular momentum from the BH until the
inequality~\eqref{superradiance} is saturated. 

By plugging into Eq.~\eqref{scalareom} the ansatz
\begin{equation}\label{ansatzscalar}
\Psi(t,\textbf{r})=\frac{1}{\sqrt{2\mu}}\left[\psi(t,\textbf{r})e^{-i\mu t}+\psi^* (t,\textbf{r})e^{i\mu t}\right]\,,
\end{equation}
where $\psi(t,\textbf{r})$ is a complex scalar field that varies on timescales much longer than $\mu^{-1}$ and $^*$ denotes
complex conjugation, and keeping only terms up to first order in
$r^{-1}$ and linear in $\alpha$, we obtain
\begin{equation}\label{schrodinger}
i\frac{\partial}{\partial t}\psi(t,\textbf{r})=\left[-\frac{1}{2\mu}\nabla^2-\frac{\alpha}{r}\right]\psi(t,\textbf{r}),
\end{equation}
where $\nabla^2$ is the Laplacian operator and $\alpha\equiv M\mu$ is
the equivalent of the fine-structure constant for the hydrogen
atom. In fact, Eq.~\eqref{schrodinger} is formally equivalent to the
Schr\"odinger equation for the hydrogen atom, and thus the eigenstates
$\psi_{n\ell m}(r)$ re hydrogenic eigenfunctions with principal and
orbital quantum numbers $n$ and $\ell$, respectively. Let us remark
that we follow the conventions of~\cite{Baumann:2018vus}, which are
more convenient to highlight similarities with the spectrum of the
hydrogen atom. In particular, our principal quantum number $n$ is the
same as $\tilde n=n+\ell+1$ in Dolan's notation~\cite{Dolan:2007mj},
and the dominant superradiant mode -- a nodeless ($n=0$) solution with
$\ell=m=1$ in Dolan's notation -- corresponds to $n=2$ in our
conventions.  The eigenfrequencies of these states
are~\cite{Detweiler:1980uk}
\begin{equation}\label{eigenfreqs}
\omega_{n\ell m}\simeq \mu\left(1-\frac{\alpha^2}{2n^2}+\delta\omega_{n\ell m}\right)\,,
\end{equation}
where $\delta \omega_{n\ell m}$ denote higher-order corrections, that (up to
fifth-order in $\alpha$) are given by~\cite{Baumann:2018vus}
\begin{equation}\label{eigenfreqs_2}
\delta\omega_{n\ell m} \simeq \left(-\frac{\alpha^4}{8n^4}+\frac{(2\ell-3n+1)\alpha^4}{n^4(\ell+1/2)}+\frac{2\tilde{a}m\alpha^5}{n^3\ell(\ell+1/2)(\ell+1)}\right)\,.
\end{equation}
Finally, the characteristic Bohr radius -- i.e., the radius at which
the radial profile of the scalar field achieves its maximum value --
is well approximated by
\begin{equation}
r_{\rm Bohr}\simeq \left(\frac{n^2}{\alpha^2}\right)M\,.
\end{equation} 
The eigenstates are not stationary because of dissipation, therefore
the eigenfrequencies have an imaginary part of the form
$i\Gamma_{n\ell m}$, where the coefficients $\Gamma_{n\ell m}$ are the
instability rates (decay widths) for unstable (stable) modes,
respectively~\cite{Detweiler:1980uk,Dolan:2007mj}.

For generic values of the constant $\alpha$, the decay width
$\Gamma_{n\ell m}$ must be computed numerically, as explained in
Appendix~\ref{app:continuedfrac}.
However, as first shown by Detweiler~\cite{Detweiler:1980uk}, in the
limit $\alpha \ll 1$ these quantities can be computed analytically,
with the result:
\begin{equation}\label{detweiler_width}
\Gamma_{n\ell m}=\frac{2r_+}{M}C_{n\ell m}(a,\alpha)\left(m\Omega_H-\omega\right)\alpha^{4\ell +5}\,,
\end{equation}
where
\begin{align}
C_{n\ell m}(a,\alpha)&:=\frac{2^{4\ell+1}\left(n+\ell\right)!}{n^{2\ell+4}\left(n-\ell-1\right)!}\left[\frac{\ell !}{\left(2\ell\right)!\left(2\ell+1\right)!}\right]^2\times\nn\\
&\prod^{\ell}_{j=1}\left[j^2\left(1-\tilde{a}^2\right)+\left(\tilde{a}m-2\tilde{r}_+\alpha\right)^2\right]\,.
\label{eq:Cfunction}
\end{align}

\begin{figure}%
\includegraphics[width=\columnwidth]{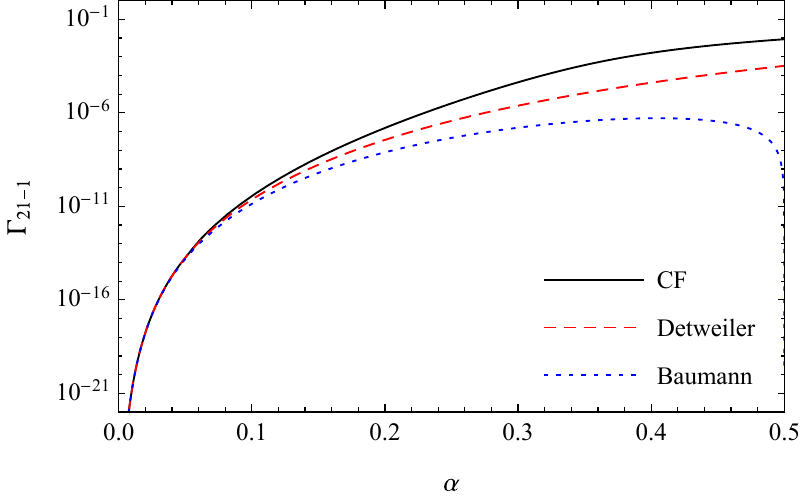}%
\caption{Decay width $\Gamma_{21-1}$ computed using continued
  fractions (solid black), the Detweiler approximation (dashed red)
  and the calculation of Ref.~\cite{Baumann:2018vus} (dotted blue), for a BH spin that saturates the superradiant amplification of the $\ell=m=1$ mode [cf. Eq.\eqref{BHspin}].}
\label{fig:Gammas}
\end{figure}

One of the main purposes of this paper is to improve over this
small-$\alpha$ approximation, which is significantly inaccurate when
$\alpha\gtrsim 0.1$. This is shown in Fig.~\ref{fig:Gammas}, where --
for illustration -- we compare the decay width of the (stable)
$\Gamma_{21-1}$ mode computed using continued fractions (CF, solid
black line) and the Detweiler small-$\alpha$ approximation (dashed red
line), for a BH spin that saturates the superradiant amplification of the $\ell=m=1$ mode [cf. Eq.\eqref{BHspin}].

When comparing our results with those of Ref.~\citep{Baumann:2018vus}
we found large deviations for $\alpha\gtrsim 0.05$. This discrepancy
is caused only in part by the use of the Detweiler approximation, and
we think that it is partly due to the use of an incorrect
equation\footnote{Eq. (3.40) of~\citep{Baumann:2018vus} reads
\begin{equation}
|\Gamma_{d}^{(i)}|=\frac{B_{(i)}}{24}\frac{\alpha^{10}}{M}\left(\frac{1-4\alpha^2}{1+4\alpha^2}\right)^2\left(\frac{2}{1+4\alpha^2}+\tilde{r}_+\right)\,.
\end{equation}
where $B_{(i)}$
is a numerical constant dependent on the transition.

From Eqs.~\eqref{detweiler_width} and~\eqref{eq:Cfunction}, the
correct result for the dominant transitions should read instead
\begin{equation}
\label{eq:detweiler2}
\Gamma_{n\ell-1}=-B_{n}\frac{\alpha ^{10} \left[16 \alpha ^4 \tilde{r}_+^2+4 \alpha ^2 \left(\tilde{r}_+^2+4 \tilde{r}_++1\right)+1\right]}{\left(4 \alpha ^2+1\right)^2}\,,
\end{equation}
where $B_{2}=1/6$ and $B_{3}=128/2187$ for the hyperfine and the Bohr
mixing, respectively. Their equation can be recovered by using $m=1$
in Eq.~\eqref{eq:Cfunction}, instead of the correct value $m=-1$.} for
the decay width $\Gamma_{21-1}$.
As shown in Fig.~\ref{fig:Gammas}, where the dotted blue line shows
the prediction from Eq.~(3.40) of~\citep{Baumann:2018vus}, the
different predictions are in good agreement for $\alpha\ll 1$, but
there are large discrepancies when $\alpha\gtrsim 0.1$.

\section{Hyperfine resonance and Bohr resonance}
\label{sec:resonances}

\subsection{Circular orbits}

At the end of the superradiant process, i.e. when the superradiant
amplification saturates, the mass and spin of the final BH are related
to the frequency of the dominant superradiant mode of the cloud
by~\cite{Brito:2014wla}
\begin{equation}\label{BHspin}
\frac{a}{M}=\frac{4mM\omega}{m^2+4(M\omega)^2}.
\end{equation}
Throughout this paper we will assume that the spin of the BH that carries the cloud is given by Eq.~\eqref{BHspin}. 

If the BH that carries the cloud is part of a binary, new cloud
instabilities arise due to the existence of resonant orbits.  The
tidal field of the companion will induce perturbations in the
potential of Eq.~\eqref{schrodinger}, which in turn induce
overlaps between different states $\psi_{n\ell m}$, also known as
level mixings. Ref.~\cite{Baumann:2018vus} studied level mixings for
binaries in quasicircular orbits with orbital frequency
\begin{equation}
\Omega=\sqrt{\frac{M+M_*}{R_*^3}},
\end{equation}
where $M_*$ and $R_*$ denote the mass of the companion and the orbital
separation, respectively. Here we briefly summarize their main
results. Defining $\Phi_*$ as the azimuthal angle of $M_*$ relative
to $M$ and setting $\Omega>0$ without loss of generality,
configurations with $\Phi_*=\Omega t$ ($\Phi_*=-\Omega t$) correspond
to orbits corotating (counterrotating) with the cloud.

Let us limit attention to binary separations greater than the Bohr
radius ($R_*>r_{\rm Bohr}$) to guarantee that the gravitational
influence between the two bodies can be analyzed using a multipole
expansion, and that corrections to the Kerr metric can be treated
perturbatively~\cite{Baumann:2018vus}. If the Bohr radius
$r_{\rm Bohr}$ is greater than the Roche radius the gravitational
attraction between the two objects will induce mass transfer from the
cloud to the companion. This critical orbital separation can be
estimated using Eggleton's fitting formula~\cite{Eggleton:1983rx}:
\begin{equation}\label{roche_radius}
  R_{*,{\rm cr}}=\left[\frac{0.49q^{-2/3}}{0.6q^{-2/3}+\ln\left(1+q^{-1/3}\right)}\right]^{-1}
  r_{\rm Bohr},
\end{equation}
where $q=M_*/M$ is the ratio between the mass $M_*$ of the companion
and the mass $M$ of the BH-cloud system. We restrict our study to the
region $R_*>\max(r_{\rm Bohr}, R_{*,{\rm cr}})$, so we can neglect mass
transfer.

Let us define $\tau_c\approx 10^9 (M/10^5M_\odot)(0.1/\alpha)^{15}$~yr
to be the boson cloud lifetime, estimated assuming that GW emission is
the only dissipative process and that other effect (such as accretion)
are negligible~\cite{Brito:2014wla}. The merger time for a binary with
orbital frequency $\Omega_0$, as estimated from the quadrupole formula
for GW emission for quasicircular orbits~\cite{Peters:1964zz}, is
\begin{equation}\label{timetomerger}
\tau_0=\frac{5}{256}\frac{M_{\rm tot}}{\nu}\left(\frac{1}{M_{\rm
      tot}\Omega_0}\right)^{8/3}\,,
\end{equation}
where $M_{\rm tot}=M+M_*$ is the total mass and
$\nu=MM_*/M_{\rm tot}^2$ is the symmetric mass ratio.  If
$\tau_c>\tau_0$, the merger occurs before the cloud is radiated
away. This relation can be translated into a bound on the initial
orbital frequency:
\begin{equation}
\Omega_0>0.042
\frac{\left(1+q\right)^{1/8}}{(M/M_\odot)q^{3/8}}\alpha^{45/8} {\rm
  Hz} \equiv \Omega_c.
	\label{eq:omegac}
\end{equation}
For circular orbits, we will only consider initial orbital frequencies
greater than the critical frequency $\Omega_c$.

The selection rules for transitions induced by the tidal potential of
the companion are discussed in Appendix~\ref{app:levelmixing}.
Considering only the dominant growing mode $\psi_{211}$, two main
resonances are of interest during the orbital evolution of the binary:

\begin{itemize}
\item[(i)] the {\em hyperfine resonance} is caused by an overlap between
$\psi_{211}$ and the decaying states $\psi_{210}$ and $\psi_{21-1}$;
\item[(ii)] the {\em Bohr resonance} is caused by an overlap between
  $\psi_{211}$ and the states $\psi_{n10}$ and $\psi_{n1-1}$, for
  $n\geq 3$.
\end{itemize}

These resonances occur when the orbital frequency $\Omega$ matches the
energy split between two states $\psi_{n\ell m}$ and
$\psi_{\tilde{n}\tilde{\ell}\tilde{m}}$, i.e. when
$\Omega\sim\Delta\omega/\Delta m = \epsilon$, where
$\Delta\omega=\omega_{n\ell
  m}-\omega_{\tilde{n}\tilde{\ell}\tilde{m}}$ and
$\Delta m=m-\tilde{m}$~\cite{Baumann:2018vus,Zhang:2018kib}. More precisely, the hyperfine and Bohr
resonances will occur when $\Omega$ matches the hyperfine splitting
$\epsilon_h$ or the Bohr splitting $|\epsilon_b|$, which -- at leading
order in $\alpha$, using Eqs.~\eqref{eigenfreqs}
and~\eqref{eigenfreqs_2} -- are given by
\begin{equation}\label{split}
  \epsilon_h=\frac{\mu}{12}\tilde{a}\alpha^5\,,
  \quad
  \epsilon_{b}^{(n)}=-\frac{n^2-4}{16n^2}\mu\alpha^2\,,
\end{equation}
where $n\geq 3$.  The hyperfine resonance will only occur for
corotating orbits (because $\epsilon_h>0$), whereas the Bohr resonance
will only occur for counterrotating orbits (because
$\epsilon_b^{(n)}<0$). For orbital frequencies outside these
resonances the mixing between the modes is perturbatively small and
can, in general, be neglected. For the Bohr mixings, we will only
consider the $n=3$ resonance: this is usually dominant because it
occurs earlier in the inspiral, and because the decay width decreases
with $n$ [cf. Eq.~\eqref{eq:detweiler2}].

Let us now define the occupation densities of the modes as $c_g(t)$
for the growing mode $\psi_{211}$, $c_d^{(h)}(t)$ for the decaying
modes of the hyperfine resonance, and $c_d^{(b)}(t)$ for the decaying
modes of the Bohr resonance. In general we have a three-level system
\be
|\psi(t) \rangle = c_g(t) |\psi_g \rangle
+ c_d^{(h)}(t)|\psi_d^{(h)} \rangle
+ c_d^{(b)}(t)|\psi_d^{(b)} \rangle\,,
\ee
where the occupation densities satisfy the normalization condition
$|c_g(t)|^2+|c_d^{(h)}(t)|^2+|c_d^{(b)}(t)|^2=1$.

Consider first the hyperfine mixing. For quasicircular, co-rotating
equatorial orbits the growing mode $\psi_{211}$ dominantly couples to
the decaying mode $\psi_{21-1}$, while the perturbative coupling to
the $\psi_{31-1}$ mode can be neglected. Therefore the occupation
densities must satisfy the normalization condition
\begin{equation}
|c_g(t)|^2+|c_d^{(h)}(t)|^2=1.
\end{equation}
Solving the perturbed Schr\"odinger equation for the coupled states (see Appendix~\ref{app:levelmixing})
 with the initial conditions $c_g(0)=1$ and $c_d^{(h)}(0)=0$
yields the following proportionality relation for the occupation
density of the decaying mode:
\begin{equation}\label{density_hyper}
|c_d^{(h)}(t)|^2=\left[1-\left(\frac{\epsilon_h\mp\Omega}{\Delta_R^{(h)}}\right)^2\right]\sin^2\left[\int_{t_0}^tdt'\Delta_R^{(h)}(t')\right],
\end{equation}
where we defined the modified Rabi frequency for the hyperfine splitting as
\begin{equation}\label{rabi_hyper}
\Delta_R^{(h)}=\sqrt{\left(9\eta\right)^2+\left(\epsilon_h\mp\Omega\right)^2},
\end{equation}
where upper sign in $\mp$ stands for the co-rotating orbits and the bottom for the counter-rotating ones, and
\begin{equation}\label{eta}
\eta=\alpha^{-3}\left(\frac{q}{R_*}\right)\left(\frac{M}{R_*}\right)^2.
\end{equation}
The phase of the oscillations in Eq.~\eqref{density_hyper} has been written as an integral over time to take into account the fact that $\Delta_R^{(h)}$ changes as the orbits shrinks due to radiation reaction~\cite{Baumann:2018vus}.

The Bohr resonance is important only for counter-rotating orbits. For equatorial orbits the mode $\psi_{310}$ decouples, and we only have to consider the decaying mode
$\psi_{31-1}$ (neglecting all modes with $n>3$). Near $\Omega\simeq |\epsilon_b|$, the phase of the
hyperfine mixing oscillates rapidly with a period of the order
$\eta^{-1}$. In this region $|c_d^{(h)}(t)|^2\sim (\eta/\epsilon_b) \ll 1$~\cite{Baumann:2018vus}, and the problem reduces again to a two-level system. Solving the perturbed
Schr\"odinger equation for the remaining states with the initial
conditions $c_g(0)=1$ and $c_d^{(b)}(0)=0$ leads to the
following occupation density for the decaying state:
\begin{equation}\label{density_bohr}
|c_d^{(b)}(t)|^2=\left[1-\left(\frac{\epsilon_b^{(3)}\mp\Omega}{\Delta_R^{(b)}}\right)^2\right]\sin^2\left[\int_{t_0}^tdt'\Delta_R^{(b)}(t')\right],
\end{equation}
where $\eta$ was given in Eq.~\eqref{eta}, and in this case the
modified Rabi frequency reads
\begin{equation}\label{rabi_bohr}
\Delta_R^{(b)}=\sqrt{\left(7.6\eta\right)^2+\left(\epsilon_b^{(3)}\mp\Omega\right)^2}.
\end{equation}

The addition of modes with $n>3$ would require, in general, the
solution of an infinite-dimensional system. We can still approximate
the system as a two-level system, as these resonances occur at
different orbital frequencies for each $n$, and in the vicinity of
each resonance only one mode dominates. However, as stated above,
higher modes are subdominant in the estimation of the depletion of the
cloud, and therefore we neglect resonances with $n>3$.

\begin{figure*}%
\includegraphics[width=\columnwidth]{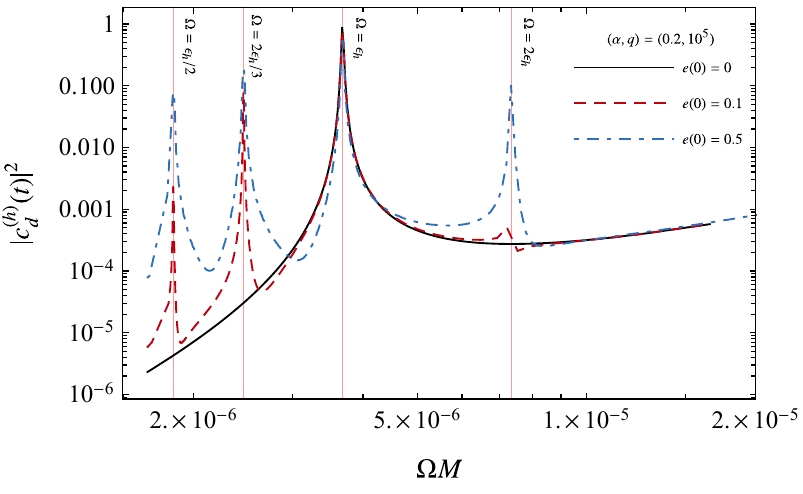}%
\includegraphics[width=\columnwidth]{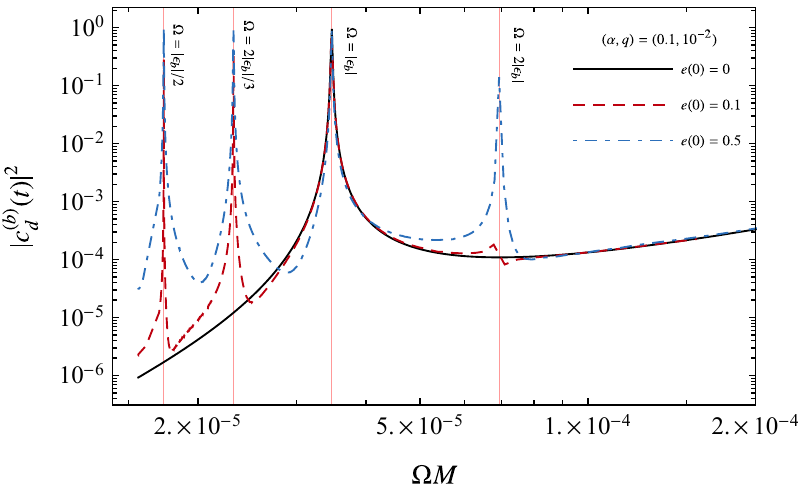}%
\caption{Left: Averaged occupation number $|c_d^{(h)}|^2$ for
  hyperfine transitions of scalar clouds in BH binaries with mass
  ratio $q=10^5$ and gravitational fine structure constant
  $\alpha=0.2$. Black lines refer to circular orbits; red, dashed
  lines and blue, dash-dotted lines refer to eccentric orbits with
  $e(0)=0.1$ and $e(0)=0.5$, respectively. For computational purposes
  we start the orbit at $\Omega(0)=0.9\epsilon_h/2$. When
  $e(0) \neq 0$, resonances can occur whenever
  $k\Omega = 2\epsilon_h$, with $k \geq 1$ an integer. The
  contribution of the different resonances to cloud depletion depends
  strongly on $e(0)$. Right: averaged occupation number
  $|c_d^{(b)}|^2$ for Bohr transitions of counterrotating orbits with
  $q=10^{-2}$ and $\alpha=0.1$.}
\label{fig:occupation_eccentric}
\end{figure*}

\subsection{Eccentric orbits}

The generalization to eccentric orbits can be done by promoting the
orbital phase $\Phi_*$ to $\Phi_*=F(\Omega t, e)$, where $\Omega$ now
describes the mean orbital frequency of the orbit and $0 \leq e < 1$
is the orbital eccentricity. At the Newtonian level, the mean orbital
frequency $\Omega$ and the orbital eccentricity $e$ are constants of
motion. To find $F(\Omega t, e)$ we use the fact that the true anomaly
$v\equiv \Phi_*-\Phi_0$ and the mean anomaly $l\equiv \Omega (t-t_0)$,
where $\Phi_0$ and $t_0$ are some initial time and initial orbital
phase, are related through the following Fourier series (see
e.g.~\cite{1961mcm..book.....B}):
\be\label{trueanom}
v = l+2\sum_{j=1}^{\infty}\frac{1}{j}\left\lbrace J_j (j e)+ \sum_{k=1}^{\infty}\beta^p
\left[J_{j-k} (j e)+J_{j+k} (j e)\right]\right\rbrace \sin{j l}\,,
\ee
where $J_j (x)$ denotes the Bessel functions of the first kind and
$\beta=(1-\sqrt{1-e^2})/e$. Without loss of generality we will set
$\Phi_0=0$ and $t_0=0$, so that $v=\Phi_*$ and $l\equiv \Omega t$.
For eccentric orbits, the binary separation will depend on the orbital
phase through the elliptical orbit equation, which at Newtonian order
is given by
\be\label{radius_ecc}
R_*=\frac{a_{\rm SM}(1-e^2)}{1+e \cos \Phi_*}\,,
\ee
where $a_{\rm SM}$ is the semi-major axis, related to the mean orbital
frequency via Kepler’s third law
\be
a_{\rm SM}=\left(\frac{M+M_*}{\Omega^2}\right)^{1/3}\,.
\ee
The cosine of the orbital phase can also be expanded in a Fourier
series (see e.g.~\cite{Yunes:2009yz}):
\be\label{cosphi}
\cos \Phi_* =-e+\frac{2}{e}(1-e^2)\sum_{j=1}^{\infty} J_j (j e) \cos{j l}\,.
\ee

As in the quasi-circular case, we only consider binary separations
greater than the critical (Roche) radius at which mass transfer from
the cloud to the companion becomes
important. Ref.~\cite{2017ApJ...844...12D} found that in the presence
of nonzero eccentricity the Roche radius in Eq.~\eqref{roche_radius}
increases by a factor $(1-e)^{-1}$. However gravitational radiation
reaction tends to circularize the orbit, and for the orbits considered
here the eccentricity is small enough that Eq.~\eqref{roche_radius} is
still a very good estimate of the Roche radius.

We now have all the necessary ingredients to generalize the
calculation of the occupation densities to eccentric orbits. As in the
circular case, let us focus on equatorial orbits and on two-state
systems, which (as argued above) describe very well the resonances of
interest. In the interaction picture, the wavefunction of the cloud is
therefore a linear combination
\be
|\psi(t) \rangle = c_g(t) |\psi_g \rangle + c_d(t)|\psi_d \rangle\,,
\ee
where $g$ and $d$ denote the growing and the decaying mode,
respectively, and again $|c_g(t)|^2+|c_d(t)|^2=1$. In the
nonrelativistic limit, and for generic equatorial orbits, the
evolution of the coefficients $\mathbf{c}\equiv (c_g,c_d)^T$ is
described by the following Schr\"{o}dinger equation [cf. Eq.~(3.21)
of~\cite{Baumann:2018vus} and Appendix~\ref{app:levelmixing}]:
\be\label{Scheq}
i \frac{d\mathbf{c}}{dt} = 
  \begin{pmatrix}
    0 & A\eta(t) e^{-i\Delta m\Phi_*(t)+i t\Delta\omega}  \\
    A\eta(t) e^{+i\Delta m\Phi_*(t)-i t \Delta\omega} & 0
  \end{pmatrix}\mathbf{c}\,,
\ee
where $A=9$ for the hyperfine resonance and $A=-7.6$ for the $n=3$
Bohr resonance described above. Below we restrict to those resonances
and define a parameter $\epsilon=\Delta\omega/\Delta m$, which is
given by Eq.~\eqref{split} for the hyperfine and Bohr transitions.

\begin{figure*}%
\includegraphics[width=\columnwidth]{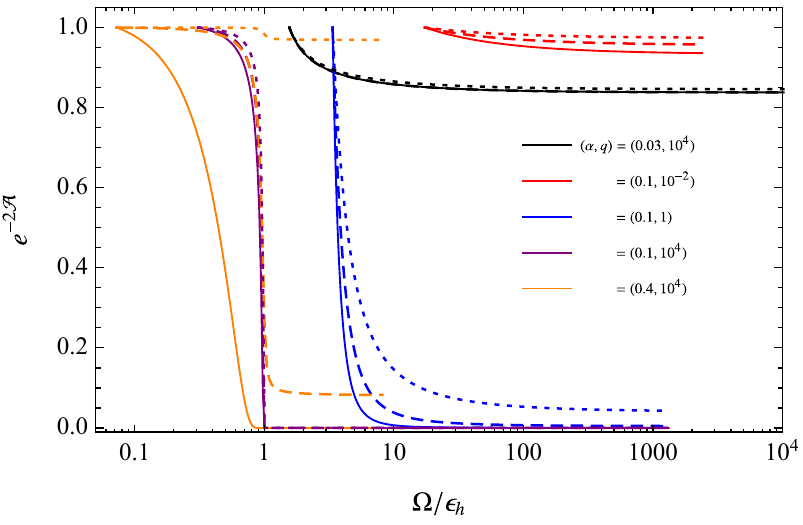}%
\includegraphics[width=\columnwidth]{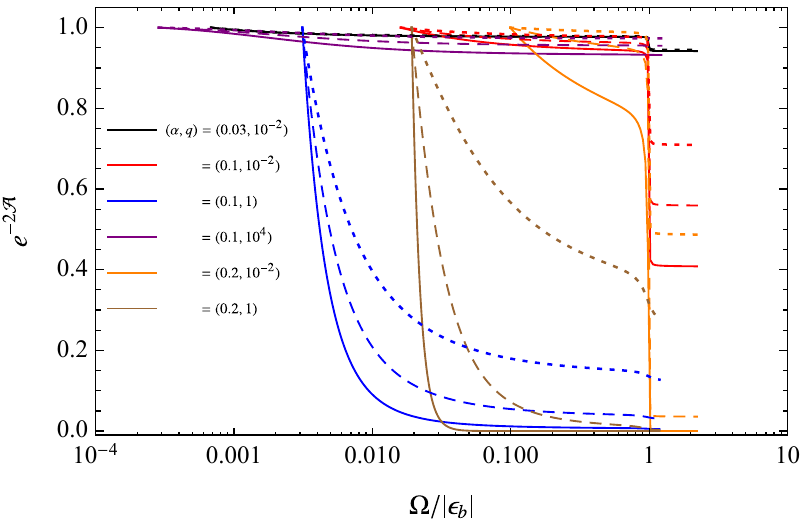}%
\caption{Left: Depletion of the scalar cloud for circular, co-rotating
  orbits (hyperfine resonances) and for selected values of $q$ and
  $\alpha$. Dotted lines correspond to results
  from~\cite{Baumann:2018vus}, dashed lines use the Detweiler
  approximation [cf.~\eqref{detweiler_width}], and solid lines use
  data generated numerically from the CF method. As expected, the
  differences are more pronounced for higher $\alpha$ (and
  particularly striking when $\alpha=0.4$). Right: same, but for
  counterrotating orbits (Bohr resonances).}%
\label{fig:coro}%
\end{figure*}

This system can be written as a single second-order differential
equation for $c_d(t)$ by taking the derivative of Eq.~\eqref{Scheq},
and by eliminating $c_g(t)$ and its derivative from the system. After
some algebra we find
\be
\ddot{c}_d=-A^2 \eta^2 c_d+\dot{c}_d\left(\frac{\dot{\eta}}{\eta}-2i(\epsilon-\dot{\Phi}_*) \right)\,,
\ee
where a dot denotes a derivative with respect to $t$. We can further simplify this equation by writing 
\be
c_d(t)=e^{-i \left[\epsilon t+B(t)\right]}C(t)\,,
\ee 
where $B(t)=B_0-\Phi_*(t)+i \log[\eta(t)]/2$ and $B_0$ is an arbitrary
integration constant. Without loss of generality we require $B(0)=0$
and therefore $B_0=\Phi_*(0)-i \log[\eta(0)]/2$. The evolution of the
system can be schematically written as
\be\label{oscillator}
\ddot{C}(t)+V(t)C(t)=0\,,
\ee
where the explicit functional form of $V(t)$ is given in
Appendix~\ref{app:ecc}.
For circular orbits $V(t)$ simplifies to
$V(t)=\Delta_R^2$, where $\Delta_R$ is the Rabi frequency defined by
\be\label{Rabi_freq}
\Delta_R=\sqrt{(A\eta_0)^2+(\epsilon\pm\Omega)^2}\,, \quad
\eta_0=\alpha^{-3}\left(\frac{q}{a_{\rm SM}}\right)\left(\frac{M}{a_{\rm SM}}\right)^2\,,
\ee
and one can easily recover the solutions first derived in
Ref.~\cite{Baumann:2018vus} and discussed above, after imposing the initial condition $C(0)=0$ and an initial condition for $\dot{C}(0)$ that can be easily found by imposing $c_g(0)=1$ in Eq.~\eqref{Scheq}.

For generic eccentricities, solutions must be found numerically for
given values of the orbital frequency $\Omega$ and of the eccentricity
$e$. Approximate analytical solutions can be found for small
eccentricity: in Appendix~\ref{app:ecc}, for illustration, we show a
solution valid up to first order in $e$.

So far we neglected gravitational radiation reaction, i.e. we assumed
that $\Omega$ and $e$ are constant. The time evolution of the mean
orbital frequency and of the eccentricity were first derived in
Ref~\cite{Peters:1964zz} under the adiabatic approximation (which is
valid when the radiation reaction time scale is much longer than the
orbital time scale). They are given by
\begin{eqnarray}
  \label{adia_equations1}
\frac{d\Omega}{dt}&=&\nu M_{\rm tot}^{5/3} \Omega^{11/3}\frac{96+292e^2+37e^4}{5(1-e^2)^{7/2}}\\
  \label{adia_equations2}
\frac{de}{dt}&=&-e\nu M_{\rm tot}^{5/3}\Omega^{8/3}\frac{304+121e^2}{15(1-e^2)^{5/2}}\,.
\end{eqnarray}

\begin{figure*}%
\includegraphics[width=\columnwidth]{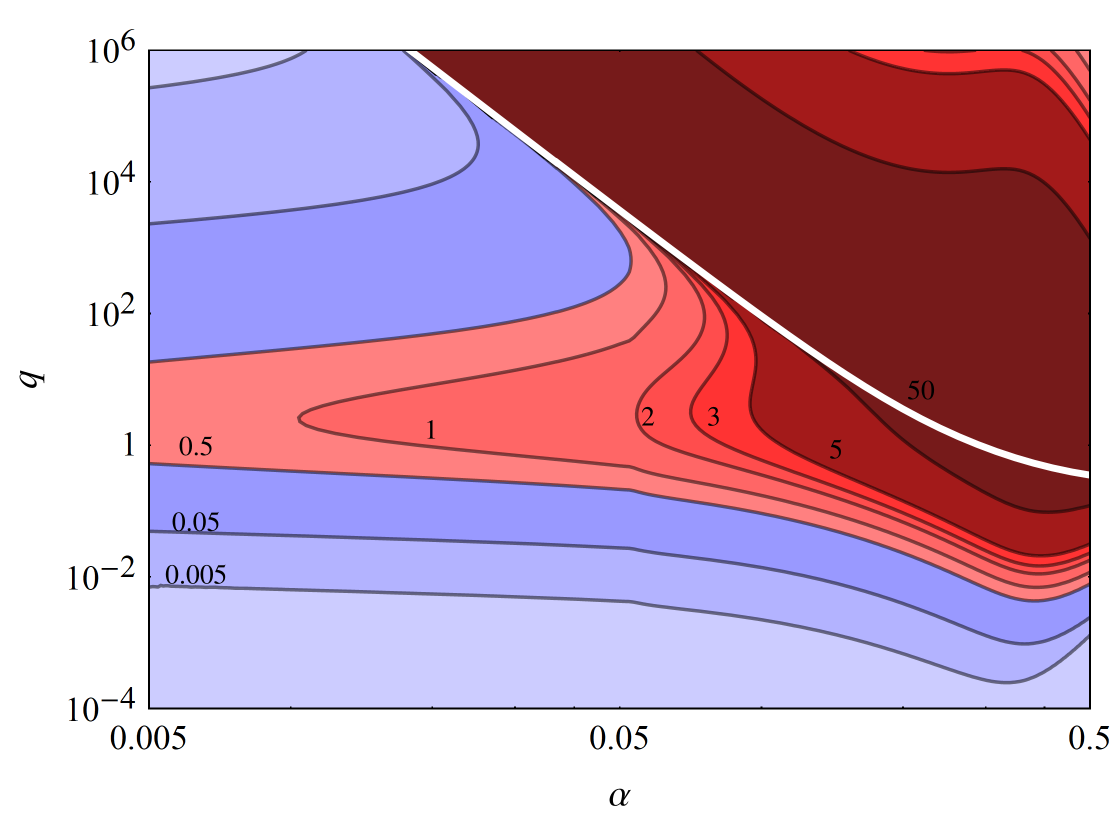}\includegraphics[width=\columnwidth]{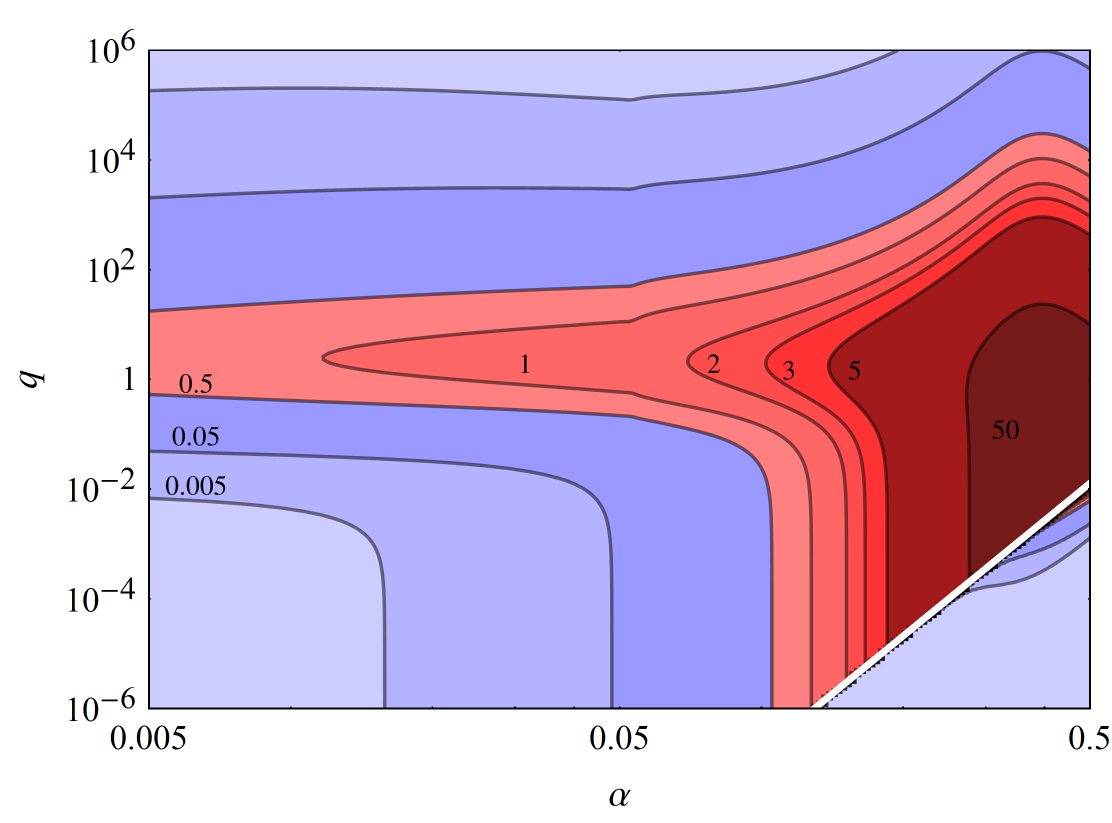}
\includegraphics[width=\columnwidth]{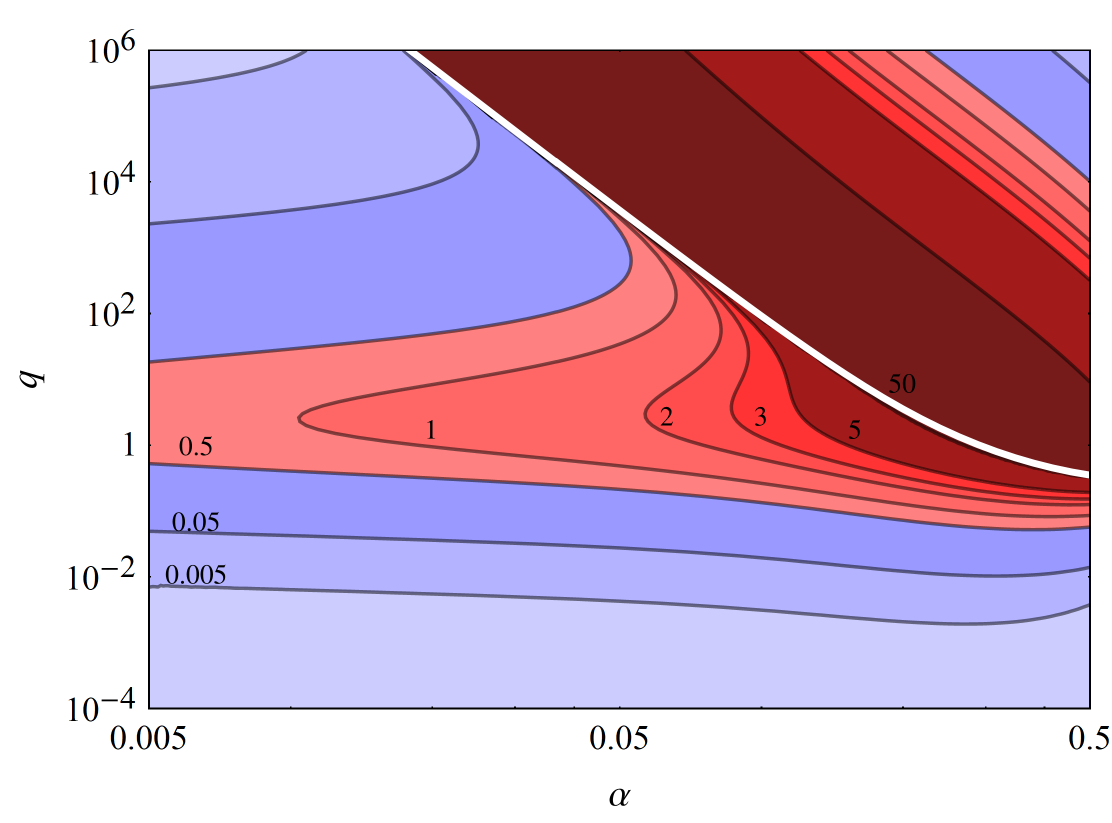}\includegraphics[width=\columnwidth]{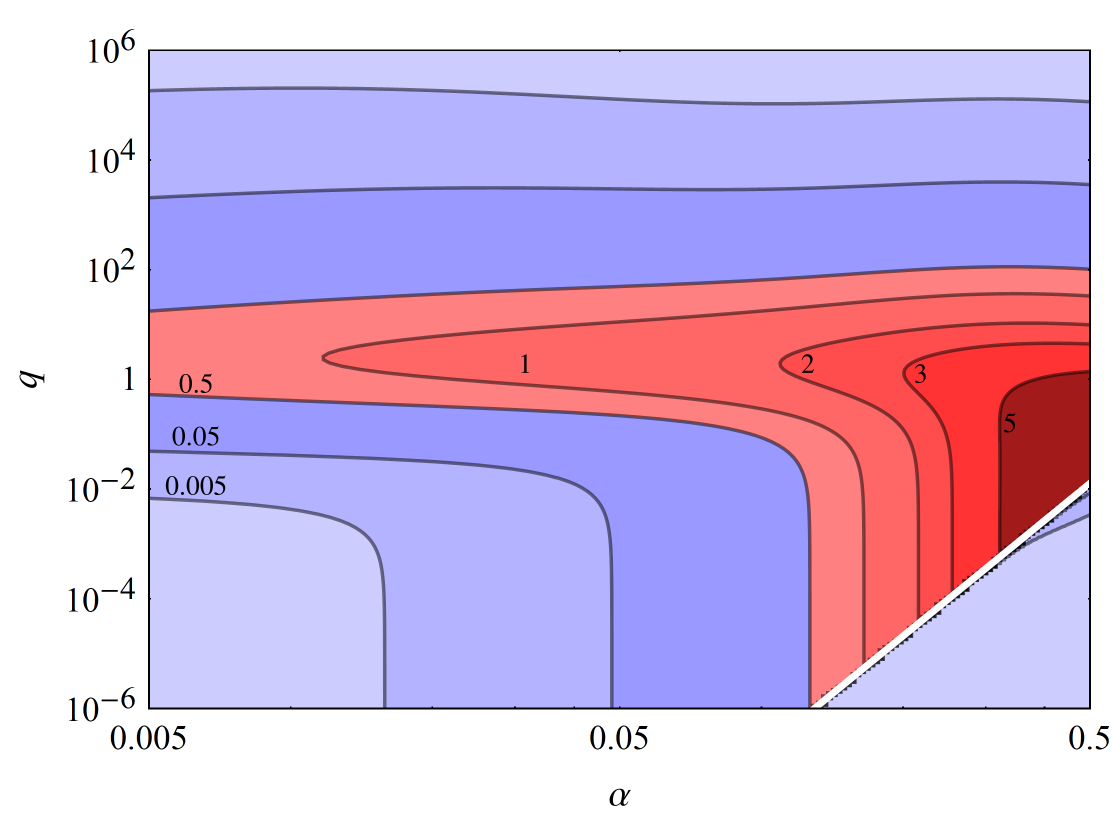}
\includegraphics[width=\columnwidth]{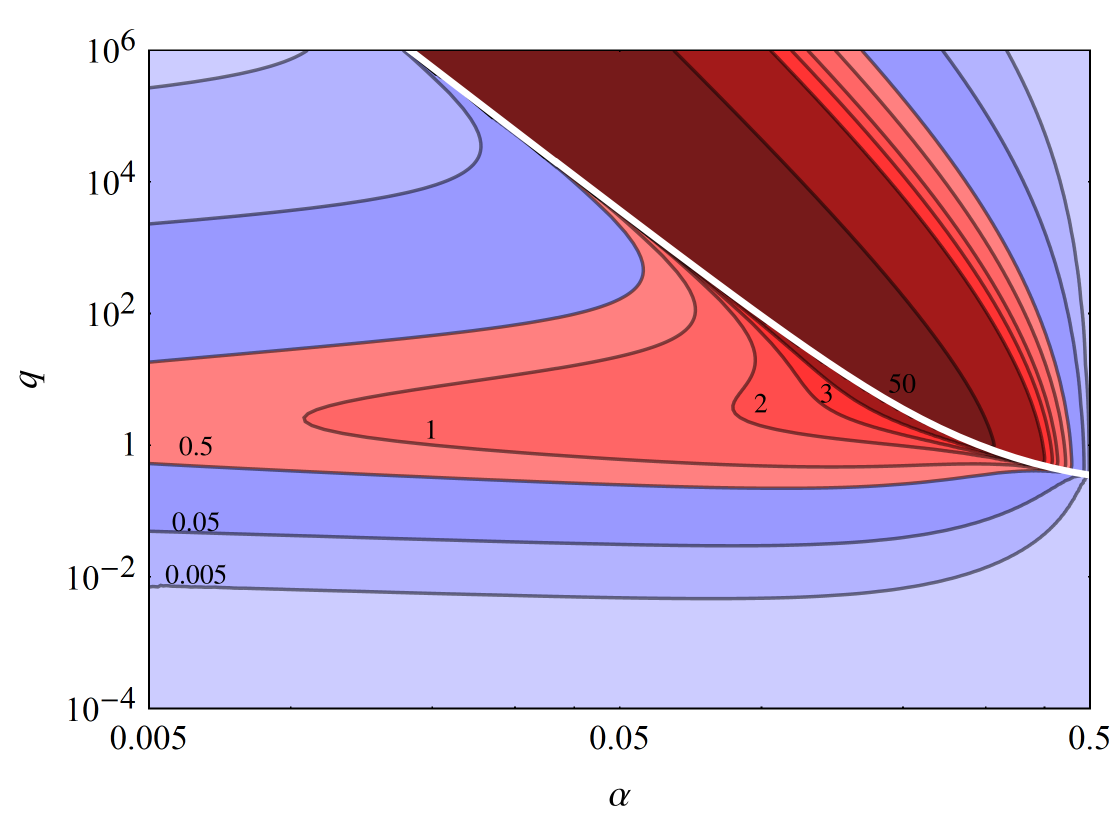}\includegraphics[width=\columnwidth]{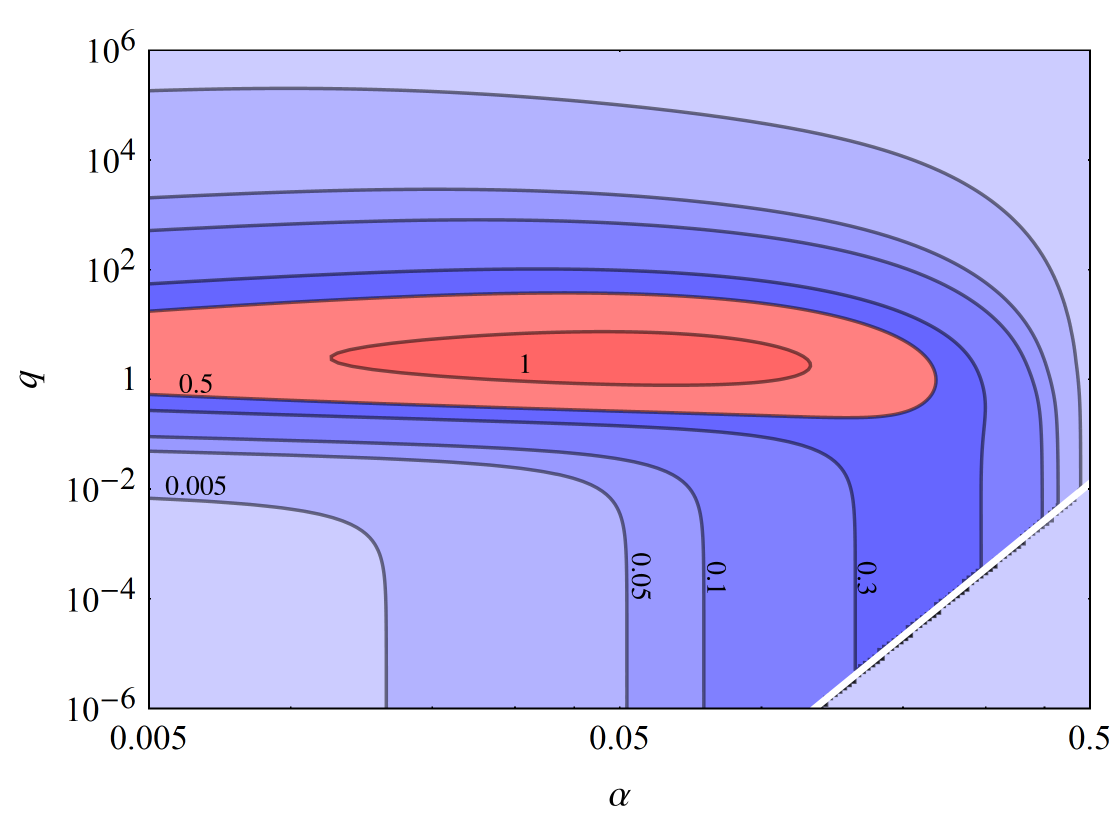}
\caption{Depletion estimator
  ${\cal A}(\Omega_{\rm f},\Omega_{c})$ for the co-rotating
  hyperfine resonances (left panels) and for the counter-rotating Bohr
  resonances (right panels). The estimators were computed using decay
  rates from the CF method (top panels) and the Detweiler
  approximation (middle panels). In the bottom row, for comparison, we
  also show results from Ref.~\cite{Baumann:2018vus}. The thick white
  lines correspond to $\Omega_c=\epsilon_h$ (left panels) or
  $\Omega_c=|\epsilon_b|$ (right panels): see the discussion in the
  main text. Recall that $q=M_*/M$, so small values of $q$ correspond
  to extreme mass-ratio inspirals of the kind discussed in
  Ref.~\cite{Hannuksela:2018izj}, while large values of $q$ correspond
  to very massive perturbing companions.}%
\label{fig:CFDetBaumann}%
\end{figure*}

To compute the equivalent of Eqs.~\eqref{density_hyper}
and~\eqref{density_bohr} for eccentric orbits, we must solve
Eq.~\eqref{oscillator} numerically while evolving the orbit
adiabatically using Eqs.~\eqref{adia_equations1} and
\eqref{adia_equations2}. The calculation of the occupation numbers can
be carried out as follows:

\begin{itemize}
\item[(i)] For a given set of fixed initial conditions
  $\Omega(t=0)=\Omega(0)$ and $e(t=0)=e(0)$, solve \eqref{oscillator}
  numerically, and average $2|c_d|^2$ over several orbits, but over
  time scales much smaller than the radiation reaction time scale;
\item[(ii)] Evolve the orbit using Eqs.~\eqref{adia_equations1} and
  \eqref{adia_equations2} with the initial conditions
  $\Omega(t=0)=\Omega(0)$ and $e(t=0)=e(0)$, and produce a grid of
  values for $e(\Omega)$.
\item[(iii)] For each value in the grid, repeat step (i) and construct
  $|c_d|^2$ as a function of $\Omega$.
\end{itemize}

For $e(0)=0$, this procedure reproduces Eqs.~\eqref{density_hyper}
and~\eqref{density_bohr} without the oscillatory terms. For
$e(0)\neq 0$, our results are shown in
Fig.~\ref{fig:occupation_eccentric}. The most important conclusion of
this calculation is that for eccentric orbits with $e(0) \neq 0$,
resonances can occur whenever $k\Omega = 2|\epsilon|$ (where
$k \geq 1$ is an integer), in contrast with the quasi-circular case
for which resonances only occur at $\Omega = |\epsilon|$.  The
existence of resonances with $k\geq 2$ implies that significant
depletion can occur {\em earlier} than in the quasi-circular case.

\begin{figure*}%
\includegraphics[width=\columnwidth]{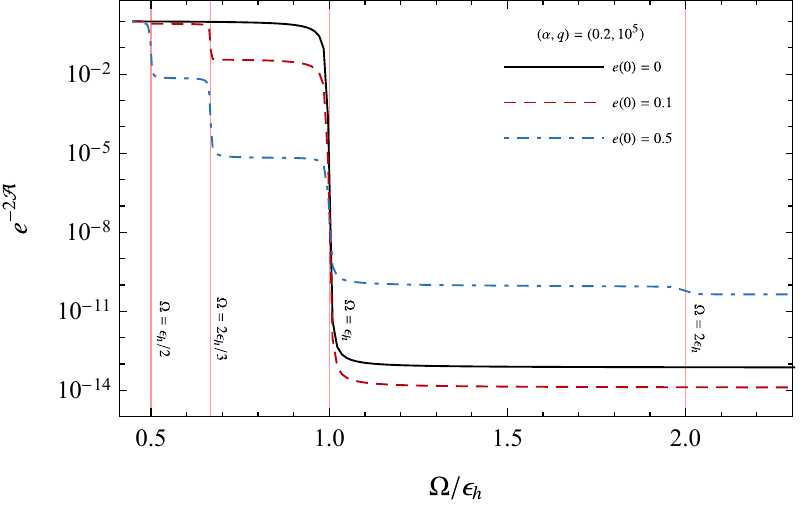}%
\includegraphics[width=\columnwidth]{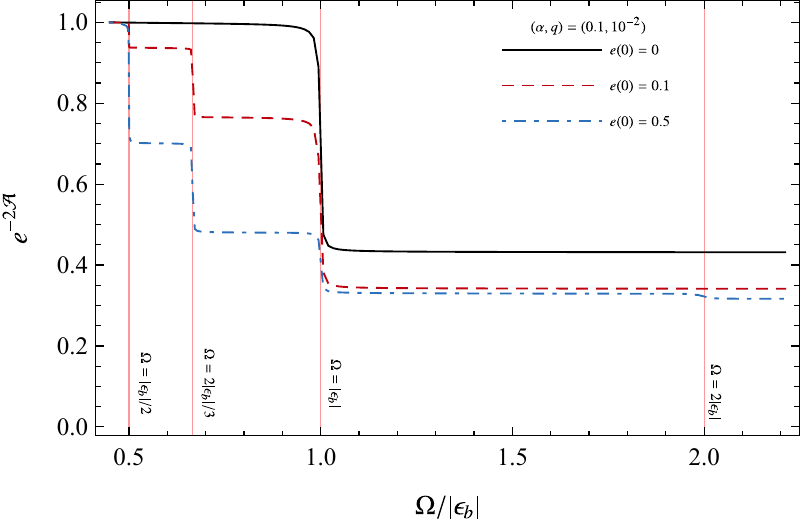}%
\caption{Depletion of the scalar cloud for two representative examples
  of eccentric binaries: a co-rotating orbit with
  $(\alpha=0.2,\, q=10^5)$ (left panel) and a counter-rotating orbit
  with $(\alpha=0.1,\, q=10^{-2})$ (right panel). The solid black,
  dashed red and dash-dotted blue lines refer to orbits with different
  initial eccentricities $e(0)=0,\,0.1$ and $0.5$, as indicated in the
  legend.}
\label{fig:depletion_eccentric}
\end{figure*}

\section{Cloud depletion: numerical results}
\label{sec:results}

Following~\cite{Baumann:2018vus}, we introduce the depletion estimator
$\mathcal A(t,\,t_0)$, where we can take $t_0$ to be the time for which superradiance
has saturated:
\begin{equation}\label{estimator}
\mathcal A(t,\,t_0)=\sum_{n,\ell}\sum_{m\leq 0}|\Gamma_{n\ell m}|\int_{t_0}^tdt'|c_{n\ell m}(t')|^2.
\end{equation}
This quantity represents the ratio between the integrated time that
the system spends in the decaying modes and the decay timescale
$|\Gamma_{n\ell m}|^{-1}$, weighted by the occupation density of each
state. The mass of the cloud decays proportionally to
$\exp\left(-2\mathcal A\right)$, where the factor of $2$ arises from
the quadratic dependence of the stress-energy tensor on the scalar
field.

The integral in Eq.~\eqref{estimator} is more easily performed in the
frequency domain.  We can make use of Eqs.~\eqref{adia_equations1} and
\eqref{adia_equations2} to write:
\begin{eqnarray}
&&\mathcal A(\Omega,\,\Omega_0)=\frac{1}{\nu M_{\rm tot}^{5/3}}\sum_{n,l}\sum_{m\leq 0}|\Gamma_{n\ell m}|\times\nonumber\\
&&\int_{\Omega_0}^{\Omega}d\Omega' \Omega'^{\,-11/3}
\frac{5(1-e^2)^{7/2}}{96+292e^2+37e^4}|c_{n\ell m}(\Omega')|^2.
\end{eqnarray}
To perform the integral we also need $e(\Omega)$. To this end we use
fits for $e(\Omega)$ from Ref.~\cite{Yunes:2009yz}, which are valid
for any initial eccentricity.  In our numerical evaluation of the
integral~\eqref{estimator} we ignore the oscillatory terms in
Eqs.~\eqref{density_hyper} and~\eqref{density_bohr}, following
Ref.~\cite{Baumann:2018vus}.

Our Figs.~\ref{fig:coro} and \ref{fig:CFDetBaumann} update the results
presented in Figs.~7 and 8 of
Ref.~\cite{Baumann:2018vus}.
To facilitate comparisons, in Fig.~\ref{fig:coro} we select the same
examples shown in the lower panels of those figures.
For quasi-circular orbits, we choose the initial frequency to be
$\Omega_0=\Omega_c$ [cf. Eq.~\eqref{eq:omegac}], and we truncate the
integral at a final frequency $\Omega$ such that our approximations
break down, i.e. $\Omega$ corresponds to an orbital radius
$R={\rm max}(r_{\rm Bohr},R_{\rm *,cr})$.

There are three possibilites, depending on the binary's mass ratio $q$
and on the gravitational fine structure constant
$\alpha$~\cite{Baumann:2018vus}:

\begin{itemize}
\item[(i)] The cloud depletes dramatically
  during the resonance.

\item[(ii)] The cloud undergoes a long period of perturbative
  depletion.

\item[(iii)] The cloud mostly survives during the entire inspiral.
\end{itemize}

Strong depletion, i.e. cases (i) and (ii), corresponds to regions
where ${\cal A}>0.5$, or $\exp(-2{\cal A})<1/e$: these regions are
marked in red in Fig.~\ref{fig:CFDetBaumann}. Most of the qualitative
features of these plots can be understood in terms of two competing
effects: (1) the decay width increases with $\alpha$; (2) for fixed
$M$ and $\alpha$, the orbital evolution due to GW emission is faster
(and hence the binary transits through resonances more rapidly) when
$q$ increases.
The thick white lines in the co-rotating (counter-rotating) cases
corresponds to setting $\Omega_c=\epsilon_h$
($\Omega_c=|\epsilon_b|$).
The binary experiences resonant depletion only when the initial
frequency $\Omega_0$ is small enough. Depletion due to co-rotating
(hyperfine) resonances occurs only in the top-right region above the
white line in the left panels of Fig.~\ref{fig:CFDetBaumann}: regions
marked in red in that part correspond to resonant depletion, i.e. case
(i) above. The cloud (partly) survives in the top-right region of
these panels only because, for very large $q$, the time that the
binary spends within the resonance is short compared to the decay time
of the decaying state. In the bottom-left region below the thick white
lines, the cloud does not experience any resonance: regions marked in
red correspond to long periods of perturbative depletion, i.e. case
(ii) above.

In contrast, depletion due to counter-rotating (Bohr) resonances
occurs almost in the entire parameter range, with the exception of a
small region in the bottom-right of the right panels of
Fig.~\ref{fig:CFDetBaumann}. 
However the Bohr resonance occurs at much
smaller orbital separations than the hyperfine resonance: for this
reason the binary moves through resonance much faster, and the amount
of depletion is (typically) significantly smaller than in the
hyperfine case.

One of the most important results of this work is that the
small-$\alpha$ approximation dramatically underestimates the decay
width $\Gamma_{21-1}$ and $\Gamma_{31-1}$ (cf. Fig.~\ref{fig:Gammas}),
and therefore it is inadequate to estimate depletion when
$\alpha\gtrsim 0.1$. This is evident in both Figs.~\ref{fig:coro} and
\ref{fig:CFDetBaumann}, with the difference being particularly
striking for the $(\alpha=0.4,\, q=10^4)$ case in the left panel of
Fig.~\ref{fig:coro}.

\begin{figure*}%
\includegraphics[width=\columnwidth]{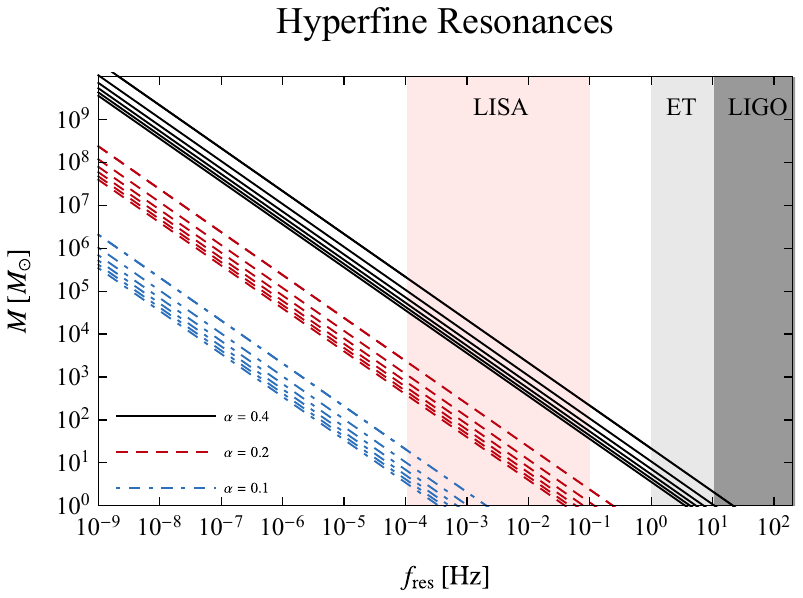}
\includegraphics[width=\columnwidth]{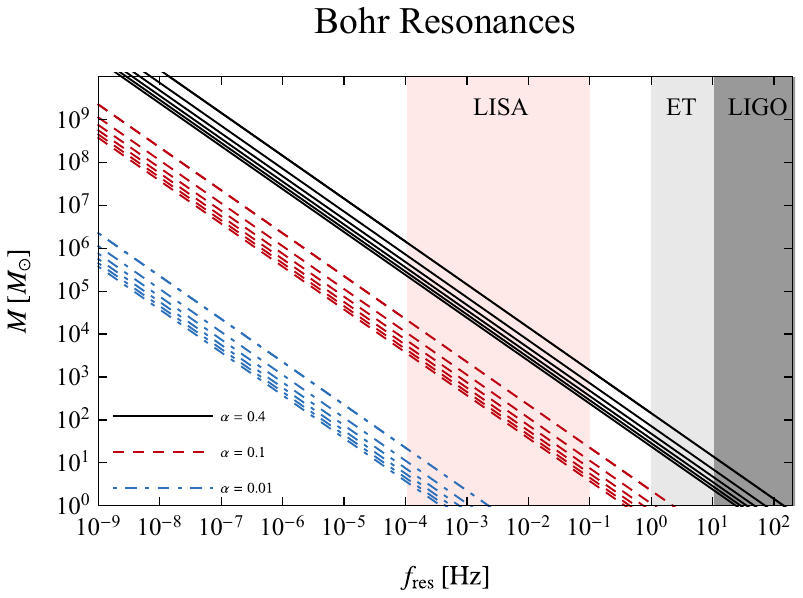}
\caption{Resonant frequencies $f_{\rm res}$ for three selected values
  of $\alpha$ ($\alpha=0.1,\,0.2,\,0.4$ for the hyperfine resonances,
  and $\alpha=0.01,\,0.1,\,0.4$ for the Bohr resonances). For each
  $\alpha$, the different lines correspond to the first six harmonics
  of the orbital frequency ($k=1,\,\dots,\,6$ from bottom to top). The
  pink shaded region corresponds to the optimal LISA sensitivity
  band. Third- (second-)generation ground-based detectors are expected
  to be sensitive for frequencies larger than a seismic cutoff
  $f\sim 1$~Hz (10~Hz), as shown by the light (dark) gray bands.}
\label{fig:frange}
\end{figure*}

Fig.~\ref{fig:depletion_eccentric} shows how cloud depletion proceeds
for the same eccentric binary systems that were shown
in Fig.~\ref{fig:occupation_eccentric}: a co-rotating orbit with
$(\alpha=0.2,\, q=10^5)$ and a counter-rotating orbit with
$(\alpha=0.1,\, q=10^{-2})$. For illustration we start the orbit at
$\Omega(0)=0.9\epsilon_h/2$ ($\Omega(0)=0.9|\epsilon_b|/2$) for the
co-rotating (counter-rotating) case. The gradual depletion at the
different resonances is apparent. The resonances with $k\geq 2$ imply
that significant depletion can occur {\em earlier} than in the
quasi-circular case, especially for large values of $e(0)$.
  
\section{Discussion and conclusions}
\label{sec:conclusion}

We reanalyzed the issue of bosonic cloud depletion in binary systems,
dropping two crucial assumptions made in Ref.~\cite{Baumann:2018vus}:
the small-$\alpha$ approximation and the assumption of circular
orbits. Our study has two important observational implications:

\begin{itemize}
\item[1)] the small-$\alpha$ approximation leads to a significant
  underestimation of cloud depletion in the region
  $\alpha\gtrsim 0.1$, where superradiant effects are expected to be
  stronger.

\item[2)] the inclusion of eccentricity significantly extends the
  frequency band in which depletion effects are potentially
  observable, because resonances can occur at all integer multiples
  $k\Omega$ of the binary's orbital frequency.
\end{itemize}

To quantify this remark, in Fig.~\ref{fig:frange} we plot resonant
frequencies -- defined as $f_{\rm res}=\epsilon/(k\pi)$ with
$\epsilon=\epsilon_h$ or $\epsilon=|\epsilon_b|$ for the hyperfine or
Bohr resonance, respectively -- for representative values of $\alpha$:
$\alpha=0.1$ roughly corresponds to the edge of the region where the
small-$\alpha$ approximation is valid, while $\alpha=0.4$ is close to
the value $\alpha=0.42$ at which the growth rate of the dominant
($\ell=m=1$) superradiant mode is maximized for near-extremal Kerr
BHs~\cite{Dolan:2007mj}. For each value of $\alpha$, the different
lines correspond to the first six harmonics of the orbital frequency
($k=1,\,\dots,\,6$ from bottom to top).

For any fixed $\alpha$, the plot shows that the frequency band in
which cloud depletion could be observable is much larger if we allow
for eccentric orbits.  Second-generation ground-based detectors are
severely limited by their seismic cutoff $f\sim 10$~Hz, as shown by
the dark gray band: there is a very small chance of observing
hyperfine resonances with $\alpha\simeq 0.4$, and the prospects are
only slightly better for Bohr resonances (which however, as discussed
earlier, would produce much smaller cloud depletion). Third-generation
detectors have better chances of observing hyperfine transitions
because of their lower seismic cutoff $f\sim 1$~Hz. Observational
prospects are much brighter for LISA, where interactions with gas and
stellar environments are expected to lead to significant orbital
eccentricities~\cite{Berti:2006ew,Roedig:2011rn}.  When superradiance
is strongest ($\alpha\simeq 0.4$), hyperfine and Bohr transitions
would occur precisely in the BH mass range
$M\sim 10^2$--$10^7~M_\odot$ targeted by LISA~\cite{Klein:2015hvg}.

Our analysis made several simplifying assumptions. We described the
orbit at Newtonian order, but going beyond this approximation is
certainly necessary for GW observations. Already at first
post-Newtonian order, orbital eccentricity adds an important physical
effect that was neglected in our work: periastron precession, which
could potentially lead to more orbital resonances. Following
Ref.~\cite{Baumann:2018vus}, we also made several assumptions (e.g. we
neglected backreaction, we treated the tidal field in a weak-field
approximation, and we assumed that there is no mass transfer) that
will inevitably break down at small orbital separations. Full
numerical simulations will be necessary to fully understand the
dynamics of the system in this regime. Other effects that should be
accounted for include self-interactions, which may lead to the
suppression of superradiant instabilities and/or
bosenovas~\cite{Arvanitaki:2010sy,Yoshino:2012kn,Yoshino:2015nsa}; the
coupling of ultralight bosons with other fields, which may lead to
electromagnetic counterparts~\cite{Rosa:2017ury,Ikeda:2019fvj,Boskovic:2018lkj};
dynamical friction and accretion, which could sensibly alter the
orbital dynamics of the
companion~\cite{Macedo:2013qea,Barausse:2014tra,Hui:2016ltb}.
All of these effects should be taken into account when modeling
gravitational waveforms.

Last but not least, from the point of view of GW data analysis it is
particularly important to understand the behavior of the energy flux
when binaries are affected by orbital resonances of the kind discussed
in this paper. Some techniques to address this problem were developed
in the context of the excitation of neutron star quasinormal modes by
orbiting companions~\cite{Pons:2001xs} and in studies of orbital
resonances for extreme mass-ratio inspirals~\cite{Gair:2010iv}. We
plan to address these issues in future work.

\section*{Acknowledgments}
We wish to thank Vitor Cardoso, Bei-Lok Hu, Lam Hui, Ted Jacobson and
Paolo Pani for discussions.
This work has received funding from the European Union’s Horizon 2020
research and innovation programme under the Marie Skłodowska-Curie
grant agreement No.~690904, and it used computational resources at the
Maryland Advanced Research Computing Center (MARCC). The authors would
like to acknowledge networking support by the GWverse COST Action
CA16104, ``Black holes, gravitational waves and fundamental physics.''
E.B. is supported by NSF Grant No.~PHY-1841464, NSF Grant
No.~AST-1841358, NSF-XSEDE Grant No.~PHY-090003, and NASA ATP Grant
No.~17-ATP17-0225.
C.F.B.M. would like to thank Conselho Nacional de Desenvolvimento
Científico e Tecnológico (CNPq), from Brazil, the
Johns Hopkins University (JHU) for kind hospitality during the early stages of
preparation of this work and the American Physical Society
which partially funded the visit to through the International Research
Travel Award Program.
R.B. acknowledges financial support from the European Union's Horizon
2020 research and innovation programme under the Marie Sk\l
odowska-Curie grant agreement No. 792862.
J.L.R. acknowledges financial support from Fundação para a Ciência e Tecnologia
(FCT) - Portugal for an FCT-IDPASC Grant No. PD/BD/114072/2015 and 
Fulbright Comission Portugal.
GR acknowledges financial support provided under the European Union's
H2020 ERC, Starting Grant agreement no.~DarkGRA--757480 and by the
H2020-MSCA-RISE-2015 Grant No. StronGrHEP-690904.

\appendix

\section{Continued-fraction method}
\label{app:continuedfrac}
The decay and growth rate of the scalar field eigenmodes can be
exactly computed by employing a continued-fraction
method~\cite{Dolan:2007mj}. For completeness, here we summarize this
method.  The Klein-Gordon equation~\eqref{scalareom} describing a
massive scalar field $\Phi$ with mass $\mu$ can be separated using the
ansatz~\eqref{scalarsep} into ordinary differential equations for the radial function
\begin{eqnarray}
&&\frac{d}{dr} \left( \Delta \frac{d \psi_{\ell m}}{d r} \right) + \left[ \frac{\omega^2 (r^2 + a^2)^2 - 4Mam\omega r + m^2 a^2 }{\Delta} \right]\psi_{\ell m}(r) \nonumber\\
&&-\left( \omega^2 a^2 + \mu^2 r^2 + \Lambda_{\ell m} \right)\psi_{\ell m}(r) = 0\,, \label{radialeq}
\end{eqnarray}
and the angular function
\begin{eqnarray}
&&\frac{1}{\sin \theta} \frac{d}{d \theta} \left(\sin \theta \frac{d S_{\ell m}}{d \theta} \right)\nonumber\\
&&+ \left[ a^2 (\omega^2 - \mu^2) \cos^2 \theta
 - \frac{m^2}{\sin^2 \theta} + \Lambda_{\ell m} \right] S_{\ell m}(\theta)  = 0\,,
\end{eqnarray}
where $\Delta=r^2+a^2-2Mr$.
When supplemented by appropriate boundary conditions, the radial and angular equations yield an eigenvalue
problem for the angular separation constant $\Lambda_{\ell m}$ and the
eigenfrequency $\omega$. The angular solutions are spheroidal harmonics with an angular separation constant that can be accurately computed through a series expansion~\cite{Berti:2005gp}
\begin{equation}\label{lambda}
\Lambda_{\ell m} = \ell (\ell + 1) + \sum_{k=1}^\infty f_{k} \, c^{2k}\,,
\end{equation}
where $c^2 = a^2 ( \omega^2 - \mu^2 )$. Explicit expressions for $f_k$
can be found in~\cite{Berti:2005gp}. Exact values for
$\Lambda_{\ell m}$ can also be computed through a continued-fraction
method~\cite{Berti:2005gp}; however, for the modes of interest
$c\ll 1$, and therefore~\eqref{lambda} provides a very accurate value
of the angular eigenvalue. Let us then focus on the calculation of the
radial eigenfrequency $\omega$.

At the event horizon the radial function goes as
\begin{equation}
\lim_{r\to r_+} \psi_{\ell m}(r) \sim (r-r_+)^{\pm i\sigma}\,,
\end{equation}
 where
$\sigma=2r_+(\omega-m\Omega_{\rm H})/(r_+-r_-)$. 
For ingoing waves at the horizon only the solution with a negative sign in the exponent is allowed. At spatial infinity the radial function behaves as
\begin{equation}
\lim_{r\to \infty} \psi_{\ell m}(r) \sim \frac{r^{(\mu^2-2\omega^2)/q}e^{qr}}{r}\,,
\end{equation}
where $q=\pm\sqrt{\mu^2-\omega^2}$. Here we will be interested in the
solutions for which $\psi_{\ell m}(r)$ is regular at infinity, and
therefore we are interested in the solutions for which
$\text{Re}(q)<0$. These solutions describe quasibound states.
 
We therefore look for solutions of the form 
\begin{equation}\label{cf_ansatz}
\psi_{\ell m}(r) = (r - r_+)^{-i \sigma} (r - r_-)^{i \sigma + \chi - 1} e^{q r} \sum_{n=0}^\infty a_n \left(\frac{r - r_+}{r - r_-}\right)^n\,,
\end{equation}
where $ \chi = (\mu^2 - 2 \omega^2)/q$ with $\text{Re}(q) < 0$. We note that choosing $\text{Re}(q) > 0$ one would instead find the quasinormal modes of the system: modes described by ingoing waves at the horizon and outgoing waves at infinity. 

After inserting this ansatz into the radial ordinary differential equation~\eqref{radialeq} one obtains a three-term recurrence relation for the coefficients $a_n$ given by
\begin{align}
\alpha_0 a_1 + \beta_0 a_0 &= 0 \\
\alpha_n a_{n+1} + \beta_n a_n + \gamma_n a_{n-1} &= 0,  \quad \quad n > 0, \quad n \in \mathbb{N},
\end{align}
where
\begin{align}
\alpha_n &= n^2 + (c_0 + 1) n + c_0 ,  \\
\beta_n   &= -2n^2 + (c_1 + 2)n + c_3 , \\
\gamma_n &= n^2 + (c_2 - 3)n + c_4 .
\end{align}
and $c_0$, $c_1$, $c_2$, $c_3$ and $c_4$ are functions of $\omega$ and $\mu$ that can be found in
Eqs.~(39)--(43) of~\cite{Dolan:2007mj}.

The ratio of the coefficients $a_n$ satisfy an infinite continued fraction
\begin{equation}
\frac{a_{n+1}}{a_n} = - \frac{\gamma_{n+1}}{\beta_{n+1}-}\frac{\alpha_{n+1}\gamma_{n+2}}{\beta_{n+2}-}\frac{\alpha_{n+2}\gamma_{n+3}}{\beta_{n+3}-} \ldots
\end{equation}
This can be further simplified after substituting the $n = 0$ term in this expression and noting that $a_1 / a_0 = - \beta_0 / \alpha_0$. We then get
\begin{equation}
\beta_0 - \frac{\alpha_0 \gamma_1}{\beta_1 -} \frac{\alpha_1 \gamma_2}{\beta_2 -} \frac{\alpha_2 \gamma_3}{\beta_3 - } \ldots = 0.
\end{equation}
The discrete family of complex values of $\omega$ for which this
condition is satisfied correspond to the bound state frequencies. In
particular for each choice of $\ell$ and $m$ there is an infinite
tower of solutions corresponding to the different overtones. In
practice, to find the solutions numerically we truncate the series
expansion in Eq.~\eqref{cf_ansatz} at some large $n$ (say, $n=10^3$)
and we empirically check that adding higher-order terms (e.g. up to
$n=10^4$) does not affect the eigenfrequencies within the desired
accuracy.

\section{Level-mixing due to gravitational perturbations}
\label{app:levelmixing}

As discussed in Ref.~\cite{Baumann:2018vus}, the gravitational
perturbations induced by a companion sufficiently far away from the
BH-cloud system can be translated into a shift in the potential of the
Schr\"{o}dinger equation~\eqref{schrodinger}, causing level
mixings. At lowest order in $\alpha$, the tidal perturbation can be written
as~\cite{Baumann:2018vus}
\begin{eqnarray}
&&V_*(t,\bar{r})=-\frac{M_*\mu}{R_*}\sum_{\ell_* \geq 2}\sum_{|m_*|\leq \ell_*}\frac{4\pi}{2\ell_*+1}\times\nonumber\\
&&\left(\frac{\bar{r}}{R_*}\right)^{\ell_*}Y^*_{\ell_*m_*}(\Theta_*,\Phi_*)Y_{\ell_*m_*}(\bar{\theta},\bar{\phi})\,,
\end{eqnarray}
where the coordinates
${\bf R}_*(t)\equiv \{R_*(t),\Theta_*(t),\Phi_*(t)\}$ describe the
position of the companion relative to the isolated BH-cloud system,
and $\bar{\bf r}\equiv \{\bar{r},\bar{\theta},\bar{\phi}\}$ are
comoving Fermi coordinates with origin at the center of mass of the
BH-cloud system.

This gravitational perturbation will induce an overlap between
different modes $\Psi_{n\ell m}$ given by
\begin{eqnarray}
&&\langle  \Psi_{j} |V_*|  \Psi_{i}\rangle = -\frac{M_*\mu}{R_*}\sum_{\ell_* \geq 2}\sum_{|m_*|\leq \ell_*}\frac{4\pi}{2\ell_*+1}\times\nonumber\\
&& \frac{Y^*_{\ell_*m_*}(\Theta_*,\Phi_*)}{R_*^{\ell}}\times I_{\bar{r}}\times I_{\bar{\Omega}}\,,
\end{eqnarray}
where
\begin{eqnarray}
I_{\bar{r}}&=&\int_0^{\infty}d\bar{r} \,\bar{r}^{2+\ell_*}\psi_{n_j\ell_j}(\bar r)\psi_{n_i\ell_i}(\bar r)\,,\\
I_{\bar{\Omega}}&=&\int d\bar{\Omega}\, Y^*_{\ell_j m_j}(\bar{\theta},\bar{\phi})Y_{\ell_i m_i}(\bar{\theta},\bar{\phi})Y_{\ell_*m_*}(\bar{\theta},\bar{\phi})\,.
\end{eqnarray}
The angular integral vanishes unless the following selection rules are
satisfied:
\begin{itemize}
\item[(i)] $-m_j+m_i+m_*=0$;
\item[(i)] $|\ell_j-\ell_i|\leq \ell_*\leq \ell_i+\ell_j$;
\item[(iii)] $\ell_i+\ell_j+\ell_*=2p$, for
  $p \in \mathbb{Z}$.
\end{itemize}
If one considers, for example, a BH-cloud system that in isolation is
only composed of the fastest growing mode $|n\ell m\rangle=|211\rangle$,
then the dominant mixings induced by the leading-order quadrupolar
perturbations $\ell_*=2$ are with the modes $|210\rangle$,
$|21-1\rangle$, $|31-1\rangle$ and $|310\rangle$. If one further restricts to
equatorial orbits $\Theta_*=\pi/2$, the mixing with the modes
$|210\rangle$ and $|310\rangle$ is forbidden, because $\langle 211 |V_*| n10 \rangle=0$
when $\Theta_*=\pi/2$.
   
Consider, for simplicity, the mixing between a growing mode $\Psi_g$
and a decaying mode $\Psi_d$ (the addition of more modes is
straightforward~\cite{Baumann:2018vus}). Using perturbation theory one
finds that the expectation value for the Hamiltonian of the field
$\Psi$ is given by
\be
H= 
\begin{pmatrix}
    E_g & 0  \\
   0 & E_d
\end{pmatrix}+
\begin{pmatrix}
    \langle  \Psi_{g} |V_*|  \Psi_{g}\rangle & \langle  \Psi_{g} |V_*|  \Psi_{d}\rangle  \\
   \langle  \Psi_{d} |V_*|  \Psi_{g}\rangle & \langle  \Psi_{d} |V_*|  \Psi_{d}\rangle
\end{pmatrix}\,,
\ee
where $E_g$ and $E_d$ are the energy eigenvalues of growing and
decaying modes for the unperturbed BH-cloud system. The Hamiltonian
can be separated into diagonal and non-diagonal parts:
\begin{eqnarray}
&&H= H_0+H_1 =
\begin{pmatrix}
    E_g+  \langle  \Psi_{g} |V_*|  \Psi_{g} \rangle & 0  \\
   0 & E_d + \langle\Psi_{d} |V_*|  \Psi_{d}\rangle
\end{pmatrix}+\nonumber\\
&&\begin{pmatrix}
    0 & \langle  \Psi_{g} |V_*|  \Psi_{d}\rangle  \\
   \langle  \Psi_{d} |V_*|  \Psi_{g}\rangle & 0
\end{pmatrix}\,.
\end{eqnarray}%
Since $H_0$ is diagonal, the eigenstates are the same as for the
isolated BH-cloud system, but with shifted energy states due to the
nonzero expectation values $\langle \Psi_{g} |V_*| \Psi_{g}
\rangle$. Calculations are then more easily performed in the
interaction picture, where the evolution of the state $|\Psi_I\rangle$
is defined by~\cite{Baumann:2018vus}
\be\label{state_int}
|\Psi_I(t)\rangle = e^{i H_0 t}|\Psi(t)\rangle\,.
\ee
In this picture, one can define the operator $H_{1,I}$ as
\be\label{ham_int}
H_{1,I}(t) = e^{iH_0 t} H_1(t) e^{-iH_0 t},
\ee
such that the Schr\"{o}dinger equation can now be written as 
\be
i \frac{d}{dt}|\Psi_I(t)\rangle = H_{1,I}(t) |\Psi_I(t)\rangle\,.
\ee
The advantage of working in the interaction picture is that $H_{1,I}$
will, in general, be a purely nondiagonal matrix
[cf. Eq.~\eqref{Scheq}]. 

In the equations above we have assumed that $H_0$ is
time-independent. When the eccentricity is nonzero, the expectation
values
$\langle \Psi_{g,d} |V_*| \Psi_{g,d}\rangle\propto R_*(t)^{-1-\ell_*}$
vary on a timescale given by the orbital period, and therefore $H_0$
is time-dependent.  We can still work in the interaction picture by
replacing the unitary propagator $e^{i H_0 t}$ with
$e^{i \int_0^t dt' H_0(t')}$ (see e.g.~\cite{Brinkmann2018}). The
integral can be simplified by noting that $H_0(t)$ varies over the
orbital period, which is much smaller than the typical timescale
associated with Rabi oscillations, especially at the resonant
frequencies [cf. Eq.~\eqref{Rabi_freq}].  Therefore the integral can
be approximated by
$\int_0^t dt' H_0(t')\approx \langle H_0 \rangle t $, where
$\langle H_0 \rangle$ denotes a time-average over the orbital
period. In addition we assume that radiation reaction can be treated
adiabatically, because $\dot{a}_{\rm SM}/a_{\rm SM}\ll \Omega$. When
computing the Hamiltonian we are also neglecting the slow decay of the
decaying modes. This is usually appropriate since the decay widths
are, in general, much smaller than the frequency eigenvalues, i.e.
$\Gamma_{n\ell m}\ll \omega_{n\ell m}$.

\section{Occupation numbers to first-order in the eccentricity}
\label{app:ecc}

The potential $V(t)$ in Eq.~\eqref{oscillator} is given by
\begin{eqnarray}
V(t)=&&A^2\eta^2+\left[\left(\epsilon-\dot{\Phi}_*\right)^2+i\ddot{\Phi}_*^2 \right]\nonumber\\
&&+\frac{2\eta\left[2i\dot{\eta}\left(\epsilon-\dot{\Phi}_*\right)+\ddot{\eta}\right]-3\dot{\eta}^2}{4\eta^2}\,.
\end{eqnarray}
Remarkably, for small eccentricities we find that
Eq.~\eqref{oscillator} can be solved exactly by expanding $V(t)$ up to
first order in $e$. In particular, at first order in $e$ one finds
\begin{eqnarray}
&&V=\Delta_R^2+\frac{e}{2}\left[\pm2i\Omega\left(\pm\Omega-3\epsilon\right)\sin(\pm \Omega t )+\right. \nonumber \\
&&\left.\left(12\Delta_R^2-12\epsilon^2 \pm 16\epsilon\Omega-7\Omega^2\right)\cos(\pm \Omega t )  \right]+\mathcal{O}(e^2)\,,
\end{eqnarray}
where we used Eqs.~\eqref{trueanom}, ~\eqref{radius_ecc}
and~\eqref{cosphi}. At zeroth order in $e$, the solution satisfying the
initial condition $C(0)=0$ is
$C(t)=A_0\sin(\Delta_R t)+\mathcal{O}(e)$. For solutions valid up to
first order in $e$ we therefore attempt to find a solution of the form
\be\label{ansatz}
C(t)=A_0\sin(\Delta_R t)+e X(t)+\mathcal{O}(e^2)\,,
\ee
where we require that $A_0$ does not depend on $e$. Requiring the
initial condition $c_g(0)=1$ and using Eq.~\eqref{Scheq} to get
$\dot{c}_h(0)$ we find
\be
A_0= -i\frac{|A|\eta_0}{\Delta_R}\,,\quad
X(0)=0\,,\quad
\dot{X}(0)=-3i|A|\eta_0\,,
\ee
Using these initial conditions for $X(t)$ and inserting~\eqref{ansatz} in Eq.~\eqref{oscillator} we find the rather complicated expression for $X(t)$
\begin{eqnarray}
X(t)=\cos\left(\Delta_R t\right)\left[a_1+b_1\sin\left(\Omega t\right)+c_1\cos\left(\Omega t\right)\right]\nonumber\\
+\sin\left(\Delta_R t\right)\left[a_2+b_2\sin\left(\Omega t\right)+c_2\cos\left(\Omega t\right)\right]\,,
\end{eqnarray}
where 
\begin{eqnarray}
a_1&=&\frac{2i A_0\Delta_R(\pm \Omega-3\epsilon)}{\Delta_e}\,,\\
b_1&=&\frac{A_0\Delta_R\left[3\Delta_e-4(\pm \Omega-\epsilon)(\pm \Omega-3\epsilon)\right]}{\Delta_e\Omega}\,,\\
c_1&=&-\frac{2i A_0\Delta_R(\pm \Omega-3\epsilon)}{\Delta_e}\,,\\
a_2&=&\frac{2 A_0(\pm \Omega-\epsilon)(\pm \Omega-3\epsilon)}{\Delta_e}-\frac{3i|A|\eta_0}{\Delta_R}-\frac{3A_0}{2}\,,\\
b_2&=&-\frac{i A_0\Omega(\pm \Omega-3\epsilon)}{\Delta_e}\,,\\
c_2&=&\frac{2A_0(\pm \Omega-\epsilon)(\pm \Omega-3\epsilon)}{\Delta_e}-\frac{3A_0}{2}\,.
\end{eqnarray}
and we introduced the quantity
\be
\Delta_e=3(\pm\Omega-2\epsilon)(\pm\Omega-2\epsilon/3)+4(A\eta_0)^2\,.
\ee
We find that this solution is in good agreement with numerical
solutions for $e\lesssim 10^{-2}$ but the approximation breaks down
for larger eccentricities, so we do not use it for the results
presented in the main text.

\bibliography{biblio}

\end{document}